\documentclass[preprint, sort&compress]{elsarticle}
\usepackage{graphicx}
\usepackage{multirow}
\usepackage[usenames]{color}
\graphicspath{{ps}}
\usepackage{pstricks}

\usepackage{url} 
\usepackage{amsmath} 
\usepackage{amssymb}

\begin{document}
\begin{frontmatter}
  \title{Study of $CP$ violation in flavor tagged and untagged $D^0 \rightarrow K^-\pi^+$ decays }
  \author[KU]{E.~Won} 
  \author[KU]{B.~R.~Ko\corref{cor1}\fnref{fn1}}
  \ead{brko@ibs.re.kr}
  \author[Ljubljana,JSI]{B.~Golob} 
  \cortext[cor1]{Corresponding author}
  \fntext[fn1]{Now at Center for Axion and Precision Physics Research, Institute for Basic Science (IBS), Daejeon 34141}

  \address[KU]{Korea University, Seoul 02841}
  \address[Ljubljana]{Faculty of Mathematics and Physics, University of Ljubljana, 1000 Ljubljana}
  \address[JSI]{J. Stefan Institute, 1000 Ljubljana}
  
  \begin{abstract}
    We review $CP$-violating observables in flavor tagged and untagged
    $D^0 \rightarrow K^{-}\pi^{+}$ decays and evaluate the $CP$ asymmetry
    difference between the two decays. We note that this commonly
    neglected difference is not zero in principle and can be significant
    in future $B$ factory experiments. We also construct an expression to
    extract the strong phase difference between $\bar{D}^0 \rightarrow
    K^-\pi^+$ and $D^0 \rightarrow K^-\pi^+$ decays, independently of existing experimental methods.
  \end{abstract}
\end{frontmatter}

The violation of combined charge conjugation and parity ($CP$)
symmetry in the quark sector through the weak interaction was
predicted by the Cabibbo-Kobayashi-Maskawa (CKM)
mechanism~\cite{ref:ckm1,ref:ckm2} and has been experimentally
observed in $K$- and $B$-meson systems~\cite{ref:pdg}. On the other
hand, $CP$ violation ($CPV$) in charm meson system has not been
observed yet and is expected to be small within the standard model
(SM)~\cite{ref:burdman}. Therefore, search for $CPV$ in charm meson
system naturally provides a window for new physics beyond the SM.

Recent experimental $CP$ asymmetry
measurements~\cite{ref:prl_K0H,ref:prl_PHIPI,ref:prl_K0PI,ref:jhep_K0K+,ref:BaBar_D0hh,ref:Belle_D0hh,ref:prl_K0P0,ref:cdf_D0hh,ref:Belle_D0hh2}
in charm meson decays adapted $D^0\rightarrow
K^-\pi^+$ \footnote{Throughout this paper, the inclusion of the
  charge-conjugate decay mode is implied unless otherwise stated.}
(referred to as ``untagged'') and $D^{*+}\rightarrow D^0(\rightarrow
K^-\pi^+)\pi^+_{\rm soft}$ (referred to as ``tagged'' and
$\pi^{+}_{\rm{soft}}$ refers to a relatively low momentum charged
pion) decays as control samples to correct for asymmetries due to
different reconstruction efficiencies between positively and
negatively charged tracks.

According to Ref.~\cite{ref:petrov}, untagged $D^0 \rightarrow
K^-\pi^+$ reveals $CP$ asymmetry resulting from the interference
between the decays with and without $D^0$-$\bar{D}^0$ mixing even
without direct $CPV$, namely $A^{\rm
  untag}_{CP}=-\sqrt{R_D}y\sin\delta\sin\phi$, which has been
considered in
Refs.~\cite{ref:prl_K0H,ref:prl_PHIPI,ref:prl_K0PI,ref:jhep_K0K+}.\footnote{Ref.~\cite{ref:prl_K0PI}
  used untagged $D^0\rightarrow K^-\pi^+\pi^0$ decays as a control
  sample , thus the consideration was modified with relevant
  $R_D$~\cite{ref:pdg} and $\delta_{K\pi\pi}$~\cite{ref:hfag}.} Other
measurements~\cite{ref:BaBar_D0hh,ref:Belle_D0hh,ref:prl_K0P0,ref:cdf_D0hh,ref:Belle_D0hh2}
used both tagged and untagged decays with the common assumption that
the difference in $CP$ asymmetries between the two decays is zero.

Also, one of the obstacles in interpreting the experimental
$D^0$-$\bar{D}^0$ mixing measurements in the decay
$\bar{D}^0\rightarrow K^-\pi^+$ is the appearance of the phase
difference between $\bar{D}^0\rightarrow K^-\pi^+$ and $D^0\rightarrow
K^-\pi^+$ decays due to the strong interaction. A direct experimental
way to extract this strong phase difference has been utilizing a
quantum-coherent production of $D^0\bar{D}^0$ pairs from $\psi(3770)$
and their measurements of the strong phase difference are $\cos\delta
= 1.15^{+0.19+0.00}_{-0.17-0.08}$, $\sin\delta =
0.56^{+0.32+0.21}_{-0.31-0.20}$, $\delta =
(18^{+11}_{-19})^{\circ}$~\cite{ref:cleo} and $\cos\delta=1.02\pm
0.11\pm 0.06\pm 0.01$~\cite{ref:besIII}. Further precise measurements
are highly desired for interpretation of the recent experimental
observations of $D^0$-$\bar{D}^0$ mixing in the decay
$\bar{D}^0\rightarrow
K^-\pi^+$~\cite{ref:dmix_lhcb,ref:dmix_cdf,ref:dmix_belle}.

In this paper, we review $CPV$ in flavor tagged and untagged
$D^0\rightarrow K^-\pi^+$ decays, testing the validity of the
systematic consideration of $CPV$ in the untagged decays adopted in
Refs.~\cite{ref:prl_K0H,ref:prl_PHIPI,ref:prl_K0PI,ref:jhep_K0K+} and the
aforementioned assumption in $CPV$ difference between the two decays
adopted in
Refs.~\cite{ref:BaBar_D0hh,ref:Belle_D0hh,ref:prl_K0P0,ref:cdf_D0hh,ref:Belle_D0hh2}
within the two $CPV$ scenarios in $D^0$ meson system,
``$CPV$-allowed'' and ``No direct $CPV$ in DCS (doubly
Cabibbo-suppressed) decays''~\cite{ref:hfag}. We also propose a method
to extract the strong phase difference between $\bar{D}^0 \rightarrow
K^-\pi^+$  and $D^0 \rightarrow K^- \pi^+$ decays.

The time evolution of the $D^0$-$\bar{D}^0$ system can be described by
the Schr\"odinger equation~\cite{ref:pdg}
\begin{equation}
i \frac{\partial }{\partial t}
\left(
\begin{array}{c}
D^0(t) \\
\bar{D}^0(t) 
\end{array}
\right)
=
\Big( \mathbf{M} - \frac{i}{2} \mathbf{\Gamma}
\Big)
\left(
\begin{array}{c}
D^0(t) \\
\bar{D}^0(t) 
\end{array}
\right)
\label{eq:mix}
\end{equation}
where $\mathbf{M}$ and $\mathbf{\Gamma}$ are Hermitian matrices
associating with the transitions, $D^0\to D^0$ and
$D^0\to\bar{D}^0$. Our mass ($|D_{1,2}\rangle$) and flavor
($|D^0\rangle, |\bar{D}^0\rangle$) eigenstates of neutral $D$ mesons
are expressed as~\cite{ref:hfag}
\begin{equation}
|D_{1,2}\rangle = p |D^0 \rangle \mp q |\bar{D}^0 \rangle
\end{equation}
where $p$ and $q$ are complex numbers with the convention
$CP|D^0\rangle=-|\bar{D}^0\rangle$ and
$CP|\bar{D}^0\rangle=-|D^0\rangle$ under $CP$ conservation.
Note that we adopt the convention used in Ref.~\cite{ref:hfag}.
The time evolution 
of the mass eigenstate is given by $|D_i(t) \rangle
= e^{-im_i t - \frac{1}{2}\Gamma_i t}|D_i\rangle$, ($i$=1,2)
where $m_i$ and $\Gamma_i$ are the mass and 
width of $|D_i\rangle$. 
From these, one usually defines mixing 
parameters 
$
x \equiv (m_1 - m_2)/\Gamma = \Delta m/\Gamma
$
and
$
y \equiv (\Gamma_1 - \Gamma_2)/2\Gamma = \Delta \Gamma/2\Gamma
$
where $\Gamma \equiv (\Gamma_1 + \Gamma_2)/2$,
in order to describe the time evolution of the $D$ meson
system and $CP$ asymmetries conveniently.

For the study of $D^0 \rightarrow f$ decay, one defines
decay amplitude of an initially produced $D^0$/$\bar{D}^0$ into the final state
$f$/$\bar{f}$ to be $\mathcal{A}_f$/$\bar{\mathcal{A}}_{\bar{f}}$
for Cabibbo-favored (CF) decays and
$\mathcal{A}_{\bar{f}}$/$\bar{\mathcal{A}}_f$ for DCS decays,
respectively, where $f$/$\bar{f}$ stands for $K^- \pi^+$/$K^+\pi^-$
final state. Then, under direct $CP$ conservation in CF decays,
$|\mathcal{A}_f|^2=|\bar{\mathcal{A}}_{\bar{f}}|^2$, we can define
\begin{eqnarray}
\label{eq:lambda}
 \frac{q}{p}\frac{\bar{\mathcal{A}}_f}{\mathcal{A}_f}
 &\equiv&
 \sqrt{R_D(1-A_D)} R_M e^{-i(\delta-\phi)},
 \nonumber \\
 \frac{p}{q}\frac{\mathcal{A}_{\bar{f}}}{\bar{\mathcal{A}}_{\bar{f}}} 
 &\equiv&
 \sqrt{R_D(1+A_D)} R^{-1}_M e^{-i(\delta+\phi)},
\\\nonumber 
\end{eqnarray}
where $R_D$ is the ratio of DCS to CF decay rates,
$R_D=\frac{|\mathcal{A}_{\bar{f}}|^2+|\bar{\mathcal{A}}_{f}|^2}{|\mathcal{A}_{f}|^2+|\bar{\mathcal{A}}_{\bar{f}}|^2}$,
$A_D$ is the direct $CPV$ in DCS decays,
$A_D=\frac{|\mathcal{A}_{\bar{f}}|^2-|\bar{\mathcal{A}}_{f}|^2}{|\mathcal{A}_{\bar{f}}|^2+|\bar{\mathcal{A}}_{f}|^2}$,
and $\delta$ is $CP$ conserving strong phase difference between DCS
and CF decay amplitudes. $R_M$ and $\phi$ are the magnitude and
argument of $q/p$,\footnote{In general, $\phi$ includes the weak phase
  arising from the corresponding complex elements of the CKM
  matrix. In charm meson decays these are, however, to a very good
  approximation real.} where $R_M\ne 1$ indicates $CPV$ in the mixing
and $\phi\ne 0$ (nor $\phi\ne\pi$) implies that in the interference of
the mixing and decay. Hence, the relations shown in
Eq.~(\ref{eq:lambda}) satisfy the condition ``$CPV$-allowed'' in
Ref.~\cite{ref:hfag}. With the relations given in
Eq.~(\ref{eq:lambda}), we have the expressions of the decay rates up
to the second order in the mixing parameters, assuming the mixing
parameters are small ($|x|\ll 1$ and $|y|\ll 1$), to be
\begin{eqnarray}
\label{eq:decayrate}
&&\Gamma({D}^0 (t) \rightarrow f) = e^{-\Gamma t} |\mathcal{A}_{f}|^2 [1
\nonumber \\
&+&\Gamma t \sqrt{R_D(1-A_D)} R_M \{y \cos{(\delta - \phi)} + x \sin{(\delta -\phi)}\}
\nonumber \\
&+&\Gamma^2 t^2\frac{1}{4}R_D(1-A_D)R^2_M(x^2+y^2)],
\nonumber \\
&&\Gamma(\bar{D}^0 (t) \rightarrow\bar{f}) = e^{-\Gamma t} |\bar{\mathcal{A}}_{\bar{f}}|^2 [1
\nonumber \\
&+&\Gamma t \sqrt{R_D(1+A_D)} R_M^{-1}\{y \cos{(\delta + \phi)} + x  \sin{(\delta +\phi)}\} 
\nonumber \\
&+&\Gamma^2 t^2\frac{1}{4}R_D(1+A_D)R^{-2}_M(x^2+y^2)],
\nonumber \\
&&\Gamma({D}^0 (t) \rightarrow\bar{f}) = e^{-\Gamma t} |\bar{\mathcal{A}}_{\bar{f}}|^2 [R_D(1+A_D) 
\nonumber \\
&+&\Gamma t \sqrt{R_D(1+A_D)} R_M\{y \cos{(\delta + \phi)} - x \sin{(\delta +\phi)}\}
\nonumber \\
&+&\Gamma^2 t^2\frac{1}{4}R^2_M(x^2+y^2)],
\nonumber \\
&&\Gamma(\bar{D}^0 (t) \rightarrow f) = e^{-\Gamma t} |\mathcal{A}_{f}|^2 [R_D(1-A_D) 
\nonumber \\
&+&\Gamma t \sqrt{R_D(1-A_D)} R_M^{-1}\{y \cos{(\delta - \phi)} - x  \sin{(\delta -\phi)}\}
\nonumber \\
&+&\Gamma^2 t^2\frac{1}{4}R^{-2}_M(x^2+y^2)],
\end{eqnarray}
and they are our fundamental relations in the construction of various
$CP$ asymmetries described below.

The final state of the untagged decay is the sum of the CF decay
$D^0\rightarrow f$, the DCS decay $\bar{D}^0\rightarrow f$, the DCS
decay following $D^0$-$\bar{D}^0$ mixing
$D^0\rightarrow\bar{D}^0\rightarrow f$, and the CF decay following
$D^0$-$\bar{D}^0$ mixing $\bar{D}^0\rightarrow D^0\rightarrow
f$. Thus, the time-integrated decay rates for the untagged case are
\begin{eqnarray}
\Gamma^{\textrm{untag}}_f
&=&
\int^\infty_0 dt~
\bigg(\Gamma(D^0 (t) \rightarrow f) + \Gamma(\bar{D}^0 (t) \rightarrow f)\bigg),
\nonumber \\
\bar{\Gamma}^{\textrm{untag}}_{\bar{f}}
&=&
\int^\infty_0 dt~ 
\bigg(\Gamma(\bar{D}^0 (t) \rightarrow \bar{f}) + \Gamma({D}^0 (t) \rightarrow \bar{f})\bigg).
\end{eqnarray}
The $CP$ asymmetry in this case is defined as
\begin{eqnarray}
A^{\textrm{untag}}_{CP} \equiv
\frac{
\Gamma^{\textrm{untag}}_f - \bar{\Gamma}^{\textrm{untag}}_{\bar{f}}
}
{
\Gamma^{\textrm{untag}}_f + \bar{\Gamma}^{\textrm{untag}}_{\bar{f}}.
}
\label{eq:acpun}
\end{eqnarray}
Note that our definition of $A_{CP}^{\textrm{untag}}$ has an opposite
sign of the one in Ref.~\cite{ref:petrov}. The expression for $A^{\rm
  untag}_{CP}$ can be evaluated using all the relations given in
Eq.~(\ref{eq:decayrate}) :
\begin{eqnarray}
\label{eq:untag}
A^{\textrm{untag}}_{CP}=
&+&2\sqrt{R_D}y\sin\delta\sin\phi
\nonumber \\
&-& 2\sqrt{R_D}(1-R_M)x\sin\delta\cos\phi 
\nonumber \\
&-& R_D A_D 
\nonumber \\
&-& \sqrt{R_D}A_D y\cos\delta\cos\phi 
\nonumber \\
&+& (1-R_M)(x^2+y^2) 
\nonumber \\
&-& R_D(1-R_M)(x^2+y^2) 
\nonumber \\
&-& \frac{1}{2}R_D A_D(x^2+y^2), 
\end{eqnarray}
where the factor 2 of the first term is not present and other terms
are neglected in Ref.~\cite{ref:petrov}.
Using the world average $CPV$ and mixing parameters from
HFAG~\cite{ref:hfag_new}, $A^{\rm
  untag}_{CP}=(+1.27\pm3.83)\times10^{-5}$, thus the magnitude of
$A^{\rm untag}_{CP}$ can be at most $8.79\times10^{-5}$ at 95\%
confidence level (CL) for the scenario ``$CPV$-allowed''. The
magnitude of $A^{\rm untag}_{CP}$ can be neglected for the current experimental
sensitivities~\cite{ref:prl_K0H,ref:prl_PHIPI,ref:prl_K0PI,ref:jhep_K0K+},
for example $A^{D^+\to
  K^0_S\pi^+}_{CP}=(-0.363\pm0.094\pm0.063\pm0.014\pm0.016)\times10^{-2}$~\cite{ref:prl_K0PI},
where the first uncertainty is statistical, the second is systematic
mostly from the usage of the control samples, the third is systematic
due to possible $CPV$ in the control sample based on
Ref.~\cite{ref:petrov}, and the fourth is irreducible systematic due
to $K^0_L$-$K^0_S$ regeneration~\cite{ref:k0mat}. The sensitivity of
$A^{D^+\to K^0_S\pi^+}_{CP}$ at the super-$B$ factory currently under
construction~\cite{ref:superb}, however, is expected to be
$(1.3\pm0.9\pm0.9\pm1.6)\times10^{-4}$, where the third is systematic
due to possible $CPV$ in the control sample given in this paper and
others are the same as above. The $CPV$ in the untagged decays will
thus become a significant systematic source in the future $CP$
asymmetry measurements. Assuming direct $CP$ conservation $A_D=0$, the
relation $1-R_M=(y/x)\tan\phi$ holds, which not only associates $CPV$
and mixing parameters, but also relates $CPV$ in mixing and
interference~\cite{ref:icpv0,ref:icpv1,ref:icpv2}. Within the SM and
also for the case ``No direct $CPV$ in DCS decays'', $\cos\phi\cong 1$
is a good approximation in charm meson system. Hereafter the
approximation is to be applied for ``No direct $CPV$ in DCS
decays''. We note that the first and second terms in
Eq.~(\ref{eq:untag}) cancel out each other under direct $CP$
conservation, then $CP$ asymmetry in the decays is expressed as
\begin{align}
\label{eq:untag2}
A^{\textrm{untag}}_{CP} = \frac{y(x^2+y^2)\sin\phi}{x}~\footnotemark
\end{align}
which is free from the strong interaction. Using the world average
$CPV$ and mixing parameters from HFAG~\cite{ref:hfag_new},
$A^{\rm{untag}}_{CP}=(-3.98\pm7.95)\times 10^{-7}$, hence the
magnitude of $A^{\rm untag}_{CP}$ can be at most $1.96\times10^{-6}$
at 95\% CL for the scenario ``No direct $CPV$ in DCS decays''. We also
find $A^{\textrm{untag}}_{CP}=0$ with limiting the relations given in
Eq.~(\ref{eq:decayrate}) up to the first order in $x$ and $y$, which
indicates indirect $CPV$ in charm decays is approximately
universal~\cite{ref:YAY}.
\footnotetext{A small additional term $-\frac{R_D y(x^2+y^2)\sin\phi}{x}$ is omitted.} 
 
For the tagged analysis, the decay $D^{*+}\to D^0(\to f)\pi^+_{\rm
  soft}$ is the sum of the CF decay $D^0\to f$ and the DCS decay
following $D^0$-$\bar{D}^0$ mixing $D^0\rightarrow\bar{D}^0\rightarrow
f$. The time-integrated decay rates for the tagged decays are
\begin{eqnarray}
\Gamma^{\textrm{tag}}_f
= 
\int^\infty_0 dt~
\Gamma({D}^0 (t) \rightarrow f),
\nonumber \\
\bar{\Gamma}^{\textrm{tag}}_{\bar{f}}
= 
\int^\infty_0 dt~
\Gamma(\bar{D}^0 (t) \rightarrow \bar{f}). 
\end{eqnarray}
The $CP$ asymmetry in the tagged decays is defined as
\begin{eqnarray}
A^{\textrm{tag}}_{CP} \equiv
\frac{
\Gamma^{\textrm{tag}}_f - \bar{\Gamma}^{\textrm{tag}}_{\bar{f}}
}
{
\Gamma^{\textrm{tag}}_f + \bar{\Gamma}^{\textrm{tag}}_{\bar{f}}
}
\label{eq:acptrs}
\end{eqnarray}
and evaluated using the first and second relations given in
Eq.~(\ref{eq:decayrate}) :
\begin{eqnarray}
\label{eq:tag}
A^{\textrm{tag}}_{CP}=
&+&\sqrt{R_D}(y\sin\delta-x\cos\delta)\sin\phi
\nonumber \\
&-&\sqrt{R_D}(1-R_M)(y\cos\delta+x\sin\delta)\cos\phi
\nonumber \\
&-&\frac{1}{2}\sqrt{R_D}A_D(y\cos\delta+x\sin\delta)\cos\phi
\nonumber \\
&-&R_D(1-R_M)(x^2+y^2)
\nonumber \\
&-&\frac{1}{2}R_D A_D(x^2+y^2).
\end{eqnarray}
Using the world average $CPV$ and mixing parameters from
HFAG~\cite{ref:hfag_new}, $A^{\rm
  tag}_{CP}=(+0.29\pm2.83)\times10^{-5}$, hence the
magnitude of $A^{\rm tag}_{CP}$ can be at most $5.85\times10^{-5}$
at 95\% CL for the scenario ``$CPV$-allowed''. Under direct $CP$
conservation, the asymmetry in the decays can be expressed as
\begin{align}
\label{eq:tag2}
A^{\textrm{tag}}_{CP} = -\frac{\sqrt{R_D}(x^2 + y^2)\cos\delta\sin\phi}{x},\footnotemark
\end{align}
where this can be found in Ref.~\cite{ref:icpv2}. Using the world
average $CPV$ and mixing parameters from HFAG~\cite{ref:hfag_new},
$A^{\rm tag}_{CP}=(+3.91\pm7.77)\times 10^{-6}$, thus the magnitude of
$A^{\rm tag}_{CP}$ can be at most $1.91\times10^{-5}$ at 95\% CL for
the scenario ``No direct $CPV$ in DCS decays''.
\footnotetext{A small additional term $-\frac{R_D y(x^2 + y^2)\sin\phi}{x}$ is omitted.}

Having compared Eq.~(\ref{eq:untag}) with Eq.~(\ref{eq:tag}) and
Eq.~(\ref{eq:untag2}) with (\ref{eq:tag2}), in general
$A^{\textrm{tag}}_{CP}\neq A^{\textrm{untag}}_{CP}$ and the
differences between them $\Delta A_{CP}$ are $(-0.98\pm2.20)\times
10^{-5}$ for the ``$CPV$-allowed'' and $(+4.31\pm8.57)\times 10^{-6}$
for the ``No direct $CPV$ in DCS decays'', respectively.
Table~\ref{TABLE:RESULTS} summarizes $CP$-violating observables and their magnitudes from this work.
\begin{table}[htbp]
\begin{center}
\caption{Summary of $CP$-violating observables and their magnitudes.}
\label{TABLE:RESULTS}
\begin{tabular}{ccc} \hline
                     &``$CPV$-allowed''                &``No direct $CPV$ \\ 
                     &                                 &  ~~~in DCS decays'' \\ \hline
$A^{\rm untag}_{CP}$ &$(+1.27\pm3.83)\times 10^{-5}$ &$(-3.98\pm7.95)\times 10^{-7}$\\ 
$A^{\rm tag}_{CP}$   &$(+0.29\pm2.83)\times 10^{-5}$ &$(+3.91\pm7.77)\times 10^{-6}$\\ 
$\Delta A_{CP}$ &$(-0.98\pm2.20)\times 10^{-5}$ &$(+4.31\pm8.57)\times 10^{-6}$\\ \hline
\end{tabular}     
\end{center}
\end{table}
Thus, the magnitudes of $\Delta A_{CP}$ can be at most
$\mathcal{O}(10^{-4})$ and $\mathcal{O}(10^{-5})$ at 95\% CL for the
``$CPV$-allowed'' and ``No direct $CPV$ in DCS decays'',
respectively. The upper limit of $\Delta A_{CP}$,
$\mathcal{O}(10^{-4})$ can be neglected for the current experimental
sensitivities~\cite{ref:BaBar_D0hh,ref:Belle_D0hh,ref:prl_K0P0,ref:cdf_D0hh,ref:Belle_D0hh2},
for example $A^{D^0\to K^+K^-}_{CP}=(-0.32\pm0.21\pm0.09)\times
10^{-2}$~\cite{ref:Belle_D0hh2}, where the first uncertainty is
statistical and the second is systematic. The sensitivity of
$A^{D^0\to K^+K^-}_{CP}$ at the super-$B$ factory~\cite{ref:superb},
however, is expected to be $3\times 10^{-4}$ (statistical) and
$1\times 10^{-4}$ (systematic). The $CPV$ difference between the
tagged and untagged decays will thus become a significant systematic
source in the future $CP$ asymmetry measurements.

\begin{figure}[htbp]
  \begin{center}
  \includegraphics[height=0.5\textwidth,width=0.6\textwidth]{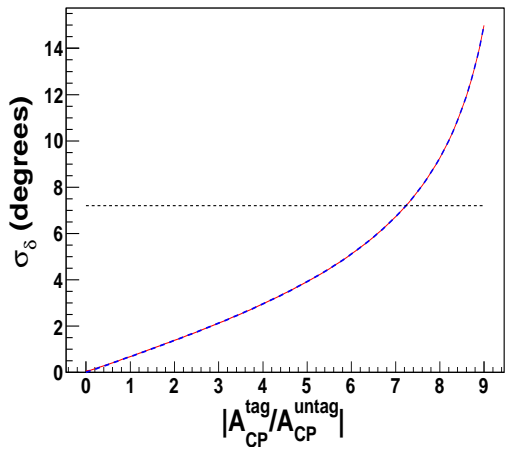}
  \caption{The solid (red) and dashed (blue) lines show the expected
    sensitivity on $\delta$ in degrees as a function of $|A^{\rm
      tag}_{CP}/A^{\rm untag}_{CP}|$ with and without experimental
    errors on $A^{\rm tag}_{CP}$ and $A^{\rm untag}_{CP}$
    measurements. The dotted (black) line is the current best
    measurement from Ref.~\cite{ref:besIII}.}
  \label{FIG:DERR}
  \end{center}
\end{figure}

For ``No direct $CPV$ in DCS decays'', the strong phase difference
between $\bar{D}^0 \rightarrow K^-\pi^+$ and $D^0\rightarrow K^-\pi^+$
decays can be obtained by taking the ratio of $A_{CP}^{\textrm{tag}}$
to $A^{\textrm{untag}}_{CP}$. The relation is
\begin{equation}
\cos{\delta}= -\frac{y}{\sqrt{R_D}}
\Bigg(
\frac{A^{\textrm{tag}}_{CP}}{A^{\textrm{untag}}_{CP}}
\Bigg),
\end{equation}
where the strong phase can be expressed in terms of $y$, $\sqrt{R_D}$,
and $CP$ asymmetries only. From this equation, one can extract the
strong phase in an independent way by measuring the ratio of $A^{\rm
  tag}_{CP}$ to $A^{\rm untag}_{CP}$ experimentally. For the
evaluation of the expected sensitivity on $\delta$
($\sigma_{\delta}$), we assign 0.007 and 0.01 for the uncertainties on
$A^{\rm untag}_{CP}$ and $A^{\rm tag}_{CP}$ measurements,
respectively, where the former is from the current best
measurement~\cite{ref:cleoIII} and the latter from our conservative
assumption reflecting a conservative experimental uncertainty. We
evaluate $\sigma_{\delta}$ as a function of $|A^{\rm tag}_{CP}/A^{\rm
  untag}_{CP}|$ by incorporating errors on $A^{\rm tag}_{CP}$, $A^{\rm
  untag}_{CP}$ given above, $y$, and $R_D$ and the correlation between
$y$ and $R_D$ from Ref.~\cite{ref:hfag_new}. Figure~\ref{FIG:DERR} shows
our evaluation implying that the sensitivity on $\delta$ using the
method introduced in this paper would be better than that of current
measurements~\cite{ref:cleo,ref:besIII} depending on $|A^{\rm
  tag}_{CP}/A^{\rm untag}_{CP}|$.
Furthermore, our evaluation shows $\sigma_{\delta}$ dominates from
current sensitivities of $y$ and $R_D$. Regardless of the sensitivity
on $\delta$, it is important to have an independent tool as a cross
check on existing methods.

To conclude, we have reviewed $CP$ asymmetries in flavor tagged and
untagged $D^0 \rightarrow K^- \pi^+$ decays, $CP$ asymmetry in the
untagged decays and $CP$ asymmetry difference between the two
decays. Both $CP$-violating observables can be ignored for the current
experimental
sensitivities~\cite{ref:prl_K0H,ref:prl_PHIPI,ref:prl_K0PI,ref:jhep_K0K+,ref:BaBar_D0hh,ref:Belle_D0hh,ref:prl_K0P0,ref:cdf_D0hh,ref:Belle_D0hh2},
but cannot be neglected in the future super-$B$ factory
experiments. However, improving $CPV$ and mixing parameters at the
future experiments helps to reduce the systematic uncertainty in the
future $CP$ asymmetry measurements. We also constructed an expression
for the strong phase difference in terms of $y$, $R_D$, and $CP$
asymmetries. This provides experimental access to the strong phase
from measurements of $CP$ asymmetries in flavor tagged and untagged
$D^0\rightarrow K^-\pi^+$ decays, which is independent of existing
methods~\cite{ref:cleo,ref:besIII}.

We thank Anze Zupanc for comments on the manuscript and Alan Schwartz
for providing us with the HFAG output correlations for $A_D=0$.
B. R. Ko is supported by the National Research Foundation of Korea (NRF) grant
funded by the Korea government (MSIP) (No. NRF-2014R1A2A2A01005286) and
E. Won is supported by No. NRF-2010-0021174.

\section{Appendix}
$CP$ asymmetry in flavor tagged $D^0\rightarrow\bar{f}$ decays
(referred to as ``wrong sign (WS)'' decays) can be expressed as
\begin{eqnarray}
A^{\textrm{tag,WS}}_{CP}=
&-&\frac{1}{\sqrt{R_D}}(y\sin\delta+x\cos\delta)\sin\phi
\nonumber \\
&-&\frac{1}{\sqrt{R_D}}(1-R_M)(y\cos\delta-x\sin\delta)\cos\phi
\nonumber \\
&+&\frac{1}{2\sqrt{R_D}}A_D(y\cos\delta-x\sin\delta)\cos\phi
\nonumber \\
&-&\frac{1}{R_D}(1-R_M)(x^2+y^2)
\nonumber \\
&+&A_D
\end{eqnarray}
using the third and fourth relations in
Eq.~(\ref{eq:decayrate}). Under direct $CP$ conservation, the
asymmetry is
\begin{eqnarray}
\label{eq:tagws}
A^{\textrm{tag,WS}}_{CP} = 
&-&\frac{(x^2 + y^2)\cos\delta\sin\phi}{\sqrt{R_D}x}
\nonumber \\
&-&\frac{y(x^2 + y^2)\sin\phi}{R_D x}.
\end{eqnarray}
Then, $A^{\rm untag}_{CP}$ in Eq.~(\ref{eq:untag2}) can be also
obtained from Eqs.~(\ref{eq:tag2}) and (\ref{eq:tagws}) because
Eq.~(\ref{eq:acpun}) can be also expressed as $A^{\rm
  untag}_{CP}=A^{\rm tag}_{CP}-R_D A^{\rm tag,WS}_{CP}$. 
By limiting the relations given in Eq.~(\ref{eq:decayrate}) up to the
first order in the mixing parameters, one can get $A^{\rm
  tag}_{CP}=R_D A^{\rm tag,WS}_{CP}$ with Eqs.~(\ref{eq:tag2}) and
(\ref{eq:tagws})~\cite{ref:icpv2}, thus $A^{\rm untag}_{CP}=0$
reflecting universality of indirect $CPV$ in charm decays in
approximation.

\end{document}